\documentclass[sigconf]{acmart}

\AtBeginDocument{%
  \providecommand\BibTeX{{%
    \normalfont B\kern-0.5em{\scshape i\kern-0.25em b}\kern-0.8em\TeX}}}

\copyrightyear{2020}
\acmYear{2020}
\setcopyright{acmlicensed}
\acmConference[AIES '20]{Proceedings of the 2020 AAAI/ACM Conference on AI, Ethics, and Society}{February 7--8, 2020}{New York, NY, USA}
\acmBooktitle{Proceedings of the 2020 AAAI/ACM Conference on AI, Ethics, and Society (AIES '20), February 7--8, 2020, New York, NY, USA}
\acmPrice{15.00}
\acmDOI{10.1145/3375627.3375872}
\acmISBN{978-1-4503-7110-0/20/02}

\begin{document}

\title{Steps Towards Value-Aligned Systems}

\author{Osonde A. Osoba}
\email{oosoba@rand.org}
\affiliation{%
  \institution{RAND Corporation}
  \streetaddress{1776 Main St}
  \city{Santa Monica}
  \state{CA}
  \postcode{90405}
}

\author{Benjamin Boudreaux}
\email{bboudreaux@rand.org}
\affiliation{%
  \institution{RAND Corporation}
  \streetaddress{1776 Main St}
  \city{Santa Monica}
  \state{CA}
  \postcode{90405}
}

\author{Douglas Yeung}
\email{dyeung@rand.org}
\affiliation{%
  \institution{RAND Corporation}
  \streetaddress{1776 Main St}
  \city{Santa Monica}
  \state{CA}
  \postcode{90405}
}

\begin{abstract}
Algorithmic (including AI/ML) decision-making artifacts are an established and growing part of our decision-making ecosystem. They are now indispensable tools that help us manage the flood of information we use to try to make effective decisions in a complex world. The current literature is full of examples of how individual artifacts violate societal norms and expectations (e.g. violations of fairness, privacy, or safety norms). Against this backdrop, this discussion highlights an under-emphasized perspective in the body of research focused on assessing value misalignment in AI-equipped sociotechnical systems. The research on value misalignment so far has a strong focus on the behavior of individual tech artifacts. This discussion argues for a more structured systems-level approach for assessing value-alignment in sociotechnical systems. We rely primarily on the research on fairness to make our arguments more concrete. And we use the opportunity to highlight how adopting a system perspective improves our ability to explain and address value misalignments better. Our discussion ends with an exploration of priority questions that demand attention if we are to assure the value alignment of whole systems, not just individual artifacts.
\end{abstract}

\begin{CCSXML}
<ccs2012>
<concept>
<concept_id>10003456.10003457.10003490.10003507.10003509</concept_id>
<concept_desc>Social and professional topics~Technology audits</concept_desc>
<concept_significance>500</concept_significance>
</concept>
<concept>
<concept_id>10003456.10003457.10003567.10010990</concept_id>
<concept_desc>Social and professional topics~Socio-technical systems</concept_desc>
<concept_significance>500</concept_significance>
</concept>
<concept>
<concept_id>10003456.10003457.10003490.10003491.10003495</concept_id>
<concept_desc>Social and professional topics~Systems analysis and design</concept_desc>
<concept_significance>300</concept_significance>
</concept>
<concept>
<concept_id>10010147.10010178.10010216</concept_id>
<concept_desc>Computing methodologies~Philosophical/theoretical foundations of artificial intelligence</concept_desc>
<concept_significance>300</concept_significance>
</concept>
<concept>
<concept_id>10010147.10010341.10010346.10010347</concept_id>
<concept_desc>Computing methodologies~Systems theory</concept_desc>
<concept_significance>300</concept_significance>
</concept>
</ccs2012>
\end{CCSXML}

\ccsdesc[500]{Social and professional topics~Technology audits}
\ccsdesc[500]{Social and professional topics~Socio-technical systems}
\ccsdesc[300]{Social and professional topics~Systems analysis and design}
\ccsdesc[300]{Computing methodologies~Philosophical/theoretical foundations of artificial intelligence}
\ccsdesc[300]{Computing methodologies~Systems theory}

\maketitle

\section{Introduction}
Issues around bias in AI/ML and algorithmic artifacts are now common knowledge.
There is a significant body of research demonstrating examples of algorithmic bias, its causes, and its features.
Typical discussions of biases inherent in AI/ML artifacts include examinations of bias in facial recognition models~\cite{buolamwini2018}, natural language processing models~\cite{caliskan2017}, and risk-estimation tools~\cite{angwin2016}.

Algorithmic bias fits into a larger body of concerns dealing with the integration and performance of technology artifacts in society.
The problem of \emph{value alignment}~\cite{arnold2017,soares2014,kim2018} captures this larger body of concerns: how can we design systems that achieve their specific objectives while remaining aligned with the broader values of society?
The values implicated can include fairness, privacy, safety, trustworthiness, etc.
The value alignment frame is useful because it primes us to think about the portfolio of values we hope to induce in our artifacts, as opposed to having an exclusionary focus on single values. This also helps us better appreciate the tensions between norms e.g. how do we balance the elevated data \emph{privacy} risks in minority sub-demographics against the need for more minority data samples for training \emph{fairer} ML models?

Much alignment-related research tends to focus narrowly on the nature of the technology artifacts and their specific value alignment pathologies.
Efforts like the MIT Moral Machine~\cite{awad2018} use crowdsourcing to elicit moral intuitions in isolated artifacts in contrived scenarios, completely stripped of the systemic perspective.
Other researchers have developed sophisticated learning mechanisms (e.g. inverse reinforcement learning models)~\cite{conitzer2017,hadfield2016} to do more than just elicit moral intuitions and try to actually induce value alignment in individual intelligent artifacts.

In contrast to these efforts, this discussion aims to draw out the implications of the fact that no technology artifact exists in a vacuum; each artifact (including AI/ML models) fulfills at least one role within a complex sociotechnical system\footnote{
We use the term \emph{sociotechnical system} (STS) to refer aggregate systems that involve complex interactions among human agents, technology artifacts, and the environment around both~\cite{baxter-sommerville2010}. For simplicity, we will focus most of our discussion on mission-oriented STSs within an institution e.g. a population of people working with diverse pieces of technologies within the criminal justice institution.
}.
Individual artifacts can have complex nonlinear interactions with the rest of the sociotechnical system. They can have hard-to-predict influences on the system's overall performance.
So even normatively perfect artifacts are not guaranteed to result in a normatively perfect system e.g. the rate of human overrides observed by Chouldechova et al.~\shortcite{chouldechova2018} after the introduction of an improved AFST tool.
There are many other ways in which the use of seemingly innocuous algorithms will be unjust in practice.
Our discussion is grounded primarily in work on fairness audits as this was the domain that informed our thinking~\cite{osoba2019}. Other relevant values for this kind of analysis include privacy and security.

Achieving value alignment in decision systems will require a broader systems-level focus in our research and implementation efforts.
And the importance of system-level value alignment audits will likely only grow as complex algorithmic decision-making artifacts become a more established part of our decision-making ecosystem.
Audits of individual artifacts will continue to be important insofar as they provide details for a more complete characterization of how decisions propagate in systems.
But perfectly aligned individual artifacts are not the goal; a fair or just sociotechnical system is the goal.

\section{A Systems Perspective in Practice?}
An important starting point for auditing value alignment in STSs is identifying the structural map of the STS.
Eliciting the exact structure of interactions in an STS can be difficult.
The goal is to gain a \emph{useful} understanding of the structure of decision-making in that system. So precision and fine detail may not be necessary.
Understanding this structure can help explain or diagnose observed systemic problems.

In practice, we have framed this elicitation as the identification of simplified \emph{decision pipelines} in an STS (e.g. in a criminal justice system, employee recruitment or promotion system, etc.). Decision pipelines are construed as collections of connected chains of decision-making points in an institution (`nodes'). Alternatively, decision pipeline can be construed as paths in an institution through which decision outcomes propagate (`edges').
The true complete structure of a typical STS is likely very complex and densely networked.

There are many advantages to this mapping exercise, not all of which need to be related to value alignment. For example, system decision maps can help:
\begin{enumerate}
  \item \textbf{Identifying Opportunities for Automated Decisions}: Identify decision points at which algorithmic decision aids may be usefully applied or are already applied. The eligibility criteria generally includes: the existence of a well-defined decision goal, the existence of significant measurable data for training ML models, \& the risks and costs associated with incorrect decisions.
  \item \textbf{Performance tracking}: Validating whole-system performance will generally involve validating the system's sub-parts. Identifying the key sub-processes will facilitate this system validation effort.
  \item \textbf{Accountability \& Explanation}: Decision system maps help focus efforts to debug failures. \emph{Credit assignment}, the ability to assign credit or blame to sub-parts of a system for observed overall errors or success, is a famously central function for effective system adaptation and optimization~\cite{minsky1961}. Credit assignment is only possible if the structure of the decision system is, at least, roughly identified.
  \item \textbf{Intervention Planning}: Decision maps also enable better planning for corrective or restorative interventions when value misalignments are in progress.
\end{enumerate}

%In this discussion, we outline and argue for a more structured systems-level approach for assessing value-alignment in sociotechnical systems.

%We use the opportunity to highlight important pathologies that come to the fore when we adopt this system perspective. Our discussion ends with an exploration of priority questions that demand attention if we are to assure the value alignment of whole systems, not just individual artifacts.

The next subsection walks through a basic sample of one such decision map developed in the course of another exercise. It is also instrumental for framing the one benefit of the systems perspective we discuss later in this paper.

\subsection{A Notional System Map: a US Criminal Justice System}
There is significant prior literature exploring the use of algorithms at different phases of the criminal justice decision pipeline.
%Risk assessment is one of the most widespread uses e.g. recidivism risk, probation violation risk, etc. This use of algorithms has also received the most outside attention.
Here we will focus on using these discussions to identify and describe a few different sub-phases of the criminal justice decision pipeline, particularly those in which there is precedent for the use of algorithmic decision aids. For example, risk assessment tools are used to determine pretrial detention decisions, sentencing, treatment/rehabilitation for incarcerated offenders, and parole/probation decisions. These tools were first used for probation decisions, then pretrial decisions, and most recently for sentencing (including some sub-decisions about treatment/rehabilitation while incarcerated).
%The statutes that authorize (and occasionally mandate) the use of these tools vary widely from state to state as some of the discussion below indicates.
Some key parts of this decision pipeline include (see Figure~\ref{cj} for a stylized depiction of the pipeline):

\begin{figure*}[!ht]
\centering
\includegraphics[width=0.75\textwidth, keepaspectratio]{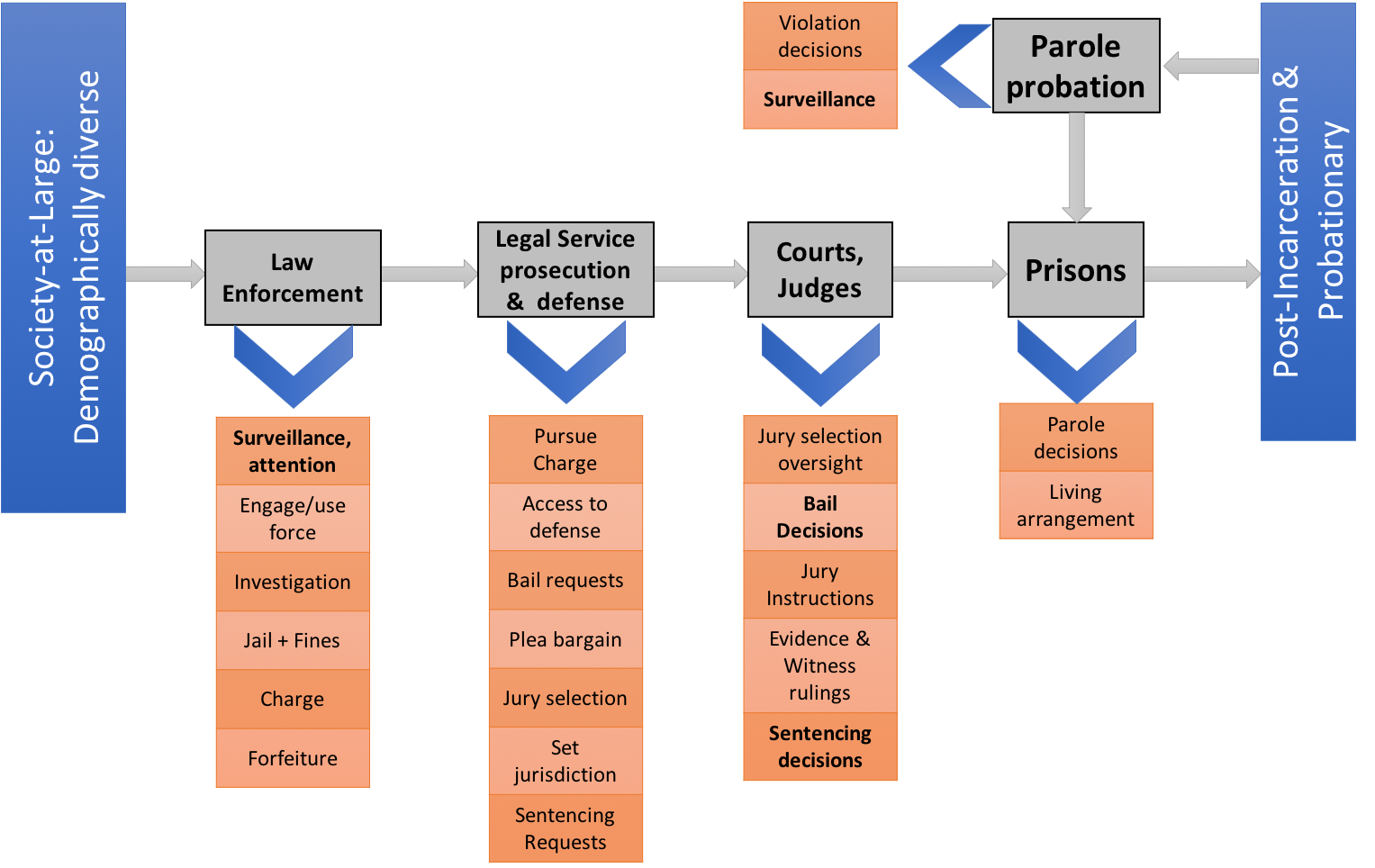} % Reduce the figure size so that it is slightly narrower than the column.
\caption{Notional example of a decision pipeline within a sociotechnical system (a US criminal justice system).}
\Description{Mapping a Criminal Justice System.}
\label{cj}
\end{figure*}

\subsubsection{Pretrial detention}
Risk-assessment tools are often used for pre-trial detention and release decisions, and generally place more focus on static risk factors in this context. One pre-trial tool, the Public Safety Assessment (PSA), is used in three states (Arizona, Kentucky, and New Jersey) and dozens of local jurisdictions in other states. The PSA was developed by the Laura and John Arnold Foundation. It was built using data from 1.5 million crimes across 300 U.S. jurisdictions in an attempt to help make decisions about whether an individual should be detained or released before trial, but not any other considerations (e.g., rehabilitation).

\subsubsection{Sentencing}
Sentencing decisions are made for convicted defendants at this decision point. Virginia was the first state to implement a recidivism risk assessment instrument for sentencing decisions relatively recently in 1994~\cite{ostrom2002}. This instrument, created by the Virginia Criminal Sentencing Commission, was designed to divert $25\%$  of the lowest-risk offenders to non-prison sanctions. In 1999, Virginia developed a second instrument; this time, the goal was to identify the highest-risk offenders.  High-risk sex offenders’ sentences could then be increased as much as threefold. In Utah, while the judge maintains discretion in the final decision, (s)he must consider both the sentencing guidelines for the offense and the recommendation of the Office of Adult Probation and Parole, which includes an assessment of risk and needs that account for factors including education level, substance abuse, and homelessness. %Using these risk assessment tools for sentencing may be difficult to implement well, as the goals for pretrial detention and probation/parole decisions are more straightforward (maintain public safety; ensure that individuals will show up for trial or to meetings with their probation officer).

%To the extent that these risk assessment instruments reflect recidivism risk, the relevance of this score to general deterrence and retribution is questionable.  Additionally, pretrial detention and parole are essentially binary decisions, and the parameters of probation are relatively circumscribed.  While there are sentencing guidelines, in most cases, judges have significant discretion in these decisions.

\subsubsection{Treatment/rehabilitation}
Some states use algorithmic scores similar to those mentioned in the sentencing section to assess which ``life skills'' classes best meet the inmate's needs and to determine conditions of probation. Newly imprisoned inmates in Pennsylvania are evaluated with the Risk Screening Tool to assess their risk of recidivism and are categorized into low, medium, and high-risk groups.  Medium and high-risk inmates may complete further assessments and potentially complete a treatment plan addressing their ``criminogenic needs'' before being considered for parole~\cite{monahan2013risk}.

\subsubsection{Parole/probation}
Rudimentary algorithms used for probability-based assessment long predate what we might consider the current algorithmic boom.  In the first half of the 20th century, Ernest W. Burgess and others created expectancy tables for parole prediction, which included factors such as work history, marital status, and prison rule violations~\cite{burgess1936}. In the 1970s, the \emph{Salient Factor Score}~\cite{hoffman1983} was first developed for use in federal parole guidelines; in that era, it was considered unfair to include immutable characteristics (e.g., race, ethnicity, gender, and age) in parole and sentencing criteria. A 1982 report also examined \emph{selective incapacitation}, a probabilistic approach to prioritizing incarceration targets by stratifying criminal risk~\cite{greenwood1982selective}.

Besides these uses, other modern decision points in the criminal justice pipeline include predictive policing, facial recognition, and probabilistic DNA analysis~\cite{cino2017}.

\section{The Fruits of a Systems-level Framing}
The value of a systems perspective on value alignment is at least two-fold: it has more explanatory power for the observed alignment pathologies; \& it gives us more tools and flexibility for correcting alignment problems. We highlight the advantages with two examples: 1) the concern around feedback cycles, and 2) an unexpected adaptation to the introduction of algorithms.

\begin{enumerate}
  \item \textbf{Algorithm overrides}: Chouldechova et al. tells of an STS pathology in a child welfare decision pipeline~\cite{chouldechova2018}. The team developed and deployed an algorithmic decision-making aid to flag high-risk cases in the child welfare system for caseworkers. The model was certified to have accuracy improvements over the status quo. However, case-workers were overriding the model's risk estimation decisions at a higher than justified rate. The reasons for the overrides were still under investigation at the time they published their results. The example illustrates that focusing on the efficiency of sub-parts of a decision-pipeline may be insufficient to improve the value-alignment of the whole system. There are \emph{impedance mismatches} or transmission barriers as information and decisions flow through an STS. This is especially true when the agents in the STS (human or algorithmic) have significant discretion in adapting to changes\footnote{The effects of such discretion or latitude are not necessarily normatively good or bad.}. Whole-system value alignment requires careful accounting for such transition costs.
  \item \textbf{Runaway feedback loops}: A myopic examination of a criminal justice decision pipeline may predispose a researcher to expect that as long as only perfectly fair algorithms are introduced the overall system will operate fairly. But simulation studies of criminal justice decision pipelines~\cite{ensign2018runaway,osoba2017intelligence} show that seemingly innocuous and fair predictive policing algorithms based on discovered crime rates (the form of data most easily available for training ML models) can lead to runaway feedback loops in inequitable patterns of criminalization (where equity is defined in terms of probability of discovery given crime commission). This particular alignment pathology can be traced to dynamics earlier in the decision pipeline, at the surveillance decision point (also further back in time). A myopic focus on just current predictive policing algorithms would be insufficient for under and addressing this misalignment.
\end{enumerate}

These documented issues make a strong case that researchers and decision-makers need to adopt a systems perspective if they are to be effective at adapting (or designing) value-aligned systems. And this is true regardless of what value, theory of justice, or philosophy the practitioner brings to the table.

\subsection{Priority Research Questions}
\begin{figure}[!ht]
\centering
\includegraphics[width=0.5\textwidth, keepaspectratio]{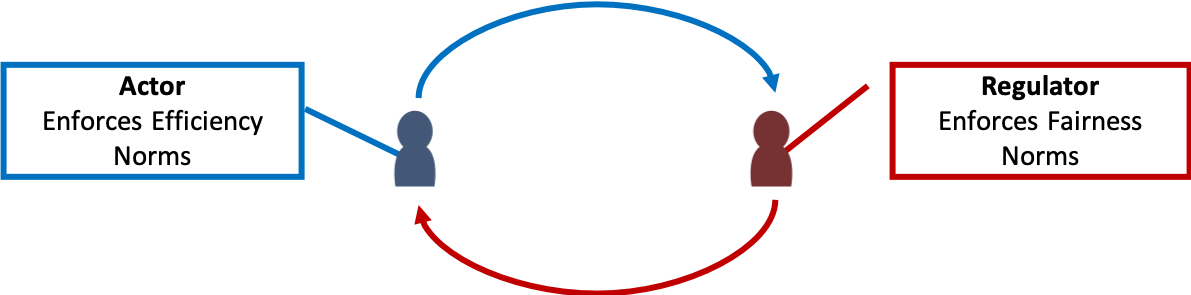}
\caption{Illustration of the interaction between free actor and governing regulator agents in an outcome-focused regulatory game. The regulator agent observes state information relevant to their regulatory mandate (in this case fairness). While the actor agent pursues its goals subject to its internal efficient mandate. Regulators may take actions to impose costs or rewards on the actors to incentivize value-aligned behavior in the population of actors.}
\Description{The Regulator-Actor Game.}
\label{acgame}
\end{figure}

A systems-level perspective is just a first step. Research programs based on this perspective (ours included) often still revert to just using system decision maps to identify individual algorithms worth auditing. Making full effective use of this framing will require a few new tools in our methodological arsenal, a minimal selection of which could include work on:

\begin{enumerate}
  \item \textbf{Characterizing Decision Transition Costs}: there can sometimes be barriers to the transmission of decisions at interfaces between AI/ML and human decision agents in a system (what we called Impedance Mismatches earlier). The reasons for such barriers can include: lack of trust in the technology, cultural aversion, the practice of human discretion, explanatory gaps (e.g. when the model is opaque), etc. There are already bodies of research on some of these concerns. Translating these types of work to better characterize and understand pathologies in sociotechnical decision systems could be enlightening.
  \item \textbf{Complex Adaptive Models of STS}: STSs are comprised of adaptive human agents/organizations and relatively static artifacts. Such complex adaptive systems (CASs) can exhibit hard-to-predict behaviors. We need principled tools and methods for modeling and simulating the interaction of human and artificial agents in these STSs. These approaches will enable better diagnostics and situational awareness for systems-level value-alignment issues.
  \item \textbf{Designing Advanced Mechanisms for Redress or Compensation and}: The focus on single artifact performance limits the range of possibilities for redress. There is a heavier focus on just designing more advanced technologies to overcome normative misalignment. This can be a myopic response. A systems perspective allows for a broader range of interventions e.g. robust oversight and compensation procedures built around imperfect artifacts.
  \item \textbf{Statistical Learning Models of Regulation}: The increasing complexity of technology artifacts (including AI/ML models) has placed a significant strain on the ability of regulating agencies to constrain the adverse impacts of these technologies. Current events routinely highlight this shortfall in regulatory capacity e.g. Boeing safety crisis of 2019, the financial crash of 2008, etc. There is a case for exploring and introducing regulatory models based on statistical learning theory and control theory, especially when the regulatory goal is focused on \emph{measurable} outcomes and consequences (as opposed to procedural norms) e.g. the regulator only cares about safety or outcome fairness. Optimal regulatory schemes in such outcome-focused or consequentialist ``regulatory games'' can be modeled as reinforcement learning problems or optimal control theory (Figure~\ref{acgame}) i.e. what optimal policies can a regulator learn to achieve a desired observable outcome in a population of agents?
  \item \textbf{Value Alignment in Multi-stakeholder STSs}: An implicit assumption in much of the work on value alignment is that \emph{the} controlling value exists and is discoverable. But recent work is beginning to highlight problems with this framing~\cite{osoba2019}. First, the assumption of a universal prescriptive controlling value for an application is very much in direct opposition with most practical experience with ethical decision-making. Second, even if such a value exists, there is no guarantee that all relevant stakeholders agree on the same controlling value(s) for an application. Resolving this conflict in a multi-stakeholder application requires resorting to preference aggregation mechanisms. There is already some sophisticated work on preference aggregation (e.g. ~\cite{sadigh2017active,conitzer2017}). There is less work targeting the elicitation of truthful preferences. And far less work on the computational mechanisms for robustly managing the process of compromise in preferences as is commonly observed in participatory modeling of stakeholder norms.
\end{enumerate}

\section{Conclusion}
In summary, this paper calls for adopting a \emph{systems engineering} approach for doing research on AI value alignment problems.
The systems perspective has advantages over the focus to-date on individual misbehaving AI/ML artifacts: it gives us more levers for redressing imperfect individual AI/ML decision artifacts; it can help us anticipate and think more clearly about mismatches that can occur as we embed more AI/ML in human decision systems; and it helps us anticipate and diagnose mechanisms that can lead to adverse feedback cycles more clearly.

Moral values and norms have been central to this discussion and its related research program. But the discussion is still agnostic as to which specific set of values are desirable to induce on  STSs and their constituent artifacts. This highlights an opportunity and a need for more philosophy and anthropology research on what broadly-acceptable binding norms to induce in technology artifacts in pluralist societies.

\bibliographystyle{ACM-Reference-Format}
% \bibliography{eqsoc,align}

\end{document}